\documentstyle[prd,aps,multicol,epsf]{revtex}
\renewcommand{\narrowtext}{\begin{multicols}{2} \global\columnwidth20.5pc}
\renewcommand{\widetext}{\end{multicols} \global\columnwidth42.5pc}
\newcommand{\Lrule}{\vspace*{-0.2in}\noindent\vrule width3.5in height.2pt
  depth.2pt \vrule depth0em height1em}
\newcommand{\Rrule}{\vspace{-0.1in}\hfill\vrule depth1em height0pt \vrule
  width3.5in height.2pt depth.2pt\vspace*{-0.1in}}
\def\beq{\begin{equation}}
\def\eeq{\end{equation}}
\def\bea{\begin{eqnarray}}
\def\eea{\end{eqnarray}}
\def\to{\rightarrow}
\def\e{{\rm e}}

\def\a{\alpha}
\begin{document}
\draft
\title{ 
Smallest Dirac Eigenvalue Distribution from Random Matrix Theory
}
\author{
Shinsuke M. Nishigaki$^{1,*}$,
Poul H. Damgaard$^{2,\dagger}$, and
Tilo Wettig$^{3,\ddagger}$}
\address{
${}^1$Institute for Theoretical Physics, University of California,
Santa Barbara, CA 93106-4030, USA}
\address{
${}^2$The Niels Bohr Institute, Blegdamsvej 17, DK-2100 Copenhagen \O,
Denmark}
\address{
${}^3$Institut f\"{u}r Theoretische Physik, Technische Universit\"{a}t
M\"{u}nchen, D-85747 Garching, Germany}
\date{March 2, 1998}
\maketitle
\begin{abstract} 
We derive the hole probability and the distribution of the smallest 
eigenvalue of chiral hermitian random matrices corresponding to  
Dirac operators coupled to massive quarks in QCD.
They are expressed in terms of the QCD partition function in the
mesoscopic regime. 
Their universality is explicitly related to that of the microscopic
massive Bessel kernel.
\end{abstract}

\narrowtext
There has long been an attractive idea that 
the low-energy physics of a complex system can be
described by a simple effective theory which respects the
global symmetries of the original system.
As an example, the quantum spectral statistics of a classically chaotic system 
is believed to be described 
by a random matrix theory belonging to the same universality class
as the former \cite{BGS}.
One new manifestation of essentially the same idea is the recent observation
that QCD Dirac operator spectra on the scale
$\lambda=O(1/V_4)$ (where $V_4$ is the space-time volume) 
measured in lattice Monte Carlo simulations \cite{BBMSVW}
are in excellent agreement with the predictions
from those large-$N$ random matrix theories \cite{V,ADMN} that
share the same global symmetries as QCD. 
The suitably rescaled (microscopic) spectral correlation functions thus 
seem to provide exact finite-size scaling functions for QCD in a
finite volume. 
Very recently, the microscopic
spectral correlators have been calculated from random matrix theories
that include the effect of fermion determinants with masses 
$m \simeq O(1/V_4)$ \cite{DN1,DN2,WGW} (see also \cite{JNZ}).
When $\lambda$ and $m$ are measured in units of
the mean level spacing at zero virtuality,
all the random matrix predictions turn out to be universal, 
i.e., insensitive to the
details of the random matrix potential \cite{ADMN,DN1,DN2}.
Although the question of whether or not QCD is included in the
same universality class cannot be answered by
demonstrating the existence of the wide range of universality within random 
matrix theories, it provides strong support for the former.

{}From the field-theoretic point of view \cite{LS} it would be most 
surprising if these observables would not also be computable solely
within the framework of finite-volume generating functionals (partition
functions) for the order parameter $\langle\bar{\psi}\psi\rangle$. 
If not, large-$N$ 
random matrix theory, which in principle is foreign to the pertinent
field theory language, would seem to be a new ingredient required to
describe the observed spectral correlators. It has recently been shown
that a description entirely in terms of finite-volume partition functions
is indeed also possible \cite{D1}. 

In order to confirm by numerical simulations that the low-lying
spectra of QCD Dirac operators can be described alternatively by
large-$N$ random matrix theories, it is in practice most convenient to
measure the distribution of the smallest eigenvalue \cite{BBMSVW} and
compare that to the random matrix prediction \cite{F,WGW}.  Since the
smallest eigenvalue distribution
largely consists of the first peak of the microscopic spectral density 
(see Fig.~1, $\zeta\equiv N \lambda$), we expect it to be universal. 
\begin{figure}
\epsfxsize=180pt
  \begin{center}
    \leavevmode
\epsfbox{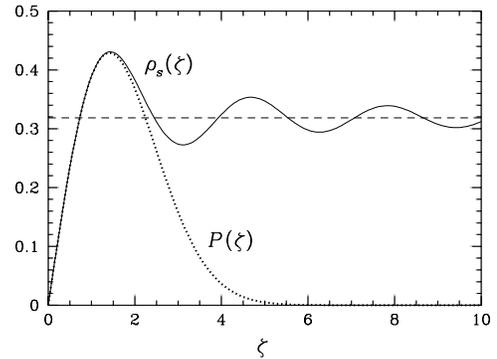}    
  \end{center}
\caption{
Microscopic spectral density (solid line,
normalized as $\rho_s(\zeta\!\to\!\infty)\!=\!1/\pi$)
and smallest eigenvalue distribution (dotted line)
for the quenched chiral unitary ensemble.
}
\end{figure}
\noindent
In fact, the proven universality of the massive kernels
\cite{DN1,DN2} of (chiral) unitary ensembles of random matrices 
guarantees the universality of the hole
probability $E(s)$, i.e., the probability that the interval $[-s,s]$
is free of eigenvalues, and of the smallest eigenvalue distribution
$P(s)=-E'(s)$ 
because it is related to 
the kernel via the Fredholm determinant formula
\cite{Meh},
\beq
E(s)
= {\rm det} (1-\hat{K}).
\label{Es}
\eeq
Here, $\hat{K}$ is an integral operator whose kernel is the 
microscopic massive Bessel kernel (Eq.~(\ref{Ks}) below
or its non-chiral counterpart) over the interval $[-s,s]$.
While other universal statistical quantities such as
the number variance $\Sigma^2(s)$
and the spectral rigidity ${\Delta}_3(s)$
of the eigenvalues in the interval $[-s,s]$
are directly related to the kernel by integral
transforms, the techniques
required to compute (\ref{Es}) from the kernel are rather involved.
In the case of the universal (massless) sine kernel describing 
the bulk of unitary ensembles, its eigenmodes are given by
the prolate spheroidal functions whose eigenvalues are known.
However, the massive generalization of these eigenmodes is 
difficult to construct.
The alternative technique of determining $E(s)$ as a  
$\tau$-function of a system of integrable
partial differential equations
\cite{TW1} does not simplify the situation either.
Consequently, the explicit calculation 
for the massive case has been done 
so far solely for the chiral Gaussian 
unitary ensemble \cite{WGW}.
The result contains an explicit factor of 
$\exp(-\zeta^2/4)$ which at first sight is due to the Gaussian potential.  
Since these results should be universal, 
the identical factor should also arise from a generic random
matrix potential.  How this happens is not immediately obvious and
will be addressed in the sequel.
We shall provide a concise method 
to circumvent the difficulty in explicitly evaluating (\ref{Es}) 
and to efficiently calculate $E(s)$
for the massive chiral unitary ensembles with generic potentials.

We define the partition function
of the chiral unitary ensemble
\beq
Z^{(\a)}(\{m\}) = \int\! dM \,
\e^{-{N} {\rm tr}\, v(M)}
\prod_{f=1}^{\a}{\det}\left(M +i  m_f\right),
\label{Z}
\eeq
where $M$ is a
$2N \times 2N$ block hermitian matrix
\beq
M=
{\small \left(
\begin{array}{cc}
0& W\\
W^\dagger &0
\end{array}
\right)} ,
\eeq
$dM$ is the Haar measure of $W$,
and $v(M)$ is an even analytic function.
The matrix $M$ models the Dirac operator for 
SU($N_c\!\geq\! 3$) QCD$_4$
in the Weyl basis, and $N$ can be identified with the spacetime volume
$V_4$.
The integer $\alpha$ corresponds to the number of flavors.  
The case of nonzero topological charge $\nu$ can be treated
by introducing $|\nu|$ massless flavors \cite{V}.  One then has
$\alpha=N_f+|\nu|$, where $N_f$ is the number of (massive) flavors and
$m_{N_f+1},\ldots,m_{N_f+|\nu|}=0$.

The (unnormalized) probability of having no eigenvalues in the interval
$[-s,s]$ is given by
\beq
E^{(\a)}(s;\{m\}) = 
\int_{|{\rm eigenvalues}|\geq s}
\!\!\!\!\!\!\!\! \!\!\!\!\!\!\!\! \!\!\!\!\!\!\!\!\! 
dM
\,\,\,\,\,\,
\e^{-{N} {\rm tr}\, v(M)}
\prod_{f=1}^{\a}{\det}\left(M +i  m_f\right) .
\eeq
It is convenient to change the picture from $M$ to
an $N\times N$ positive definite hermitian matrix
$H=W^\dagger W$,
\bea
&&E^{(\a)}(s;\{m\})
= 
\int_{{\rm eigenvalues}\geq s^2}
\!\!\!\!\!\!\!\! \!\!\!\!\!\!\!\! \!\!\!\!\!\!\!\! 
dH\,\,\,\,\,\,\,\,
\e^{-N {\rm tr}\,V(H)}
\prod_f
\det(H+m_f^2)
\nonumber\\
&&= 
\int_{{\rm eigenvalues}\geq 0}
\!\!\!\!\!\!\!\! \!\!\!\!\!\!\!\! \!\!\!\!\!\!\!\! 
dH\,\,\,\,\,\,\,\,
\e^{-N {\rm tr}\,V(H+s^2)}
\prod_f \det(H+s^2+m_f^2) ,
\label{Zs} 
\eea
where $V(z^2)\equiv 2v(z)$.
Hereafter we write
\beq
m'=\sqrt{m^2+s^2}.
\eeq
We can express (\ref{Zs}) in terms of an expectation value
$\langle\cdots\rangle_{m'}$ 
with respect to the measure
$
\e^{-N {\rm tr}\,V(H)}
\times
\prod_f \det(H+m'_f{}^2)$,
\widetext
\Lrule
\bea
E^{(\a)}(s;\{m\})
&=& 
\frac{
\int_0^\infty dH\, 
\e^{-N {\rm tr}\,V(H+s^2)}
\prod_f \det(H+m'_f{}^2)}
{
\int_0^\infty dH\, 
\e^{-N {\rm tr}\,V(H)}
\prod_f \det(H+m'_f{}^2)}
\int_0^\infty dH\, 
\e^{-N {\rm tr}\,V(H)}
\prod_f \det(H+m'_f{}^2)
\nonumber\\
&=&
\left\langle
\e^{-N {\rm tr}\,(V(H+s^2)-V(H))}
\right\rangle_{m'}
Z^{(\a)}(\{m'\}).
\label{2}
\eea
\Rrule
\narrowtext
\noindent
In the following we shall show that the two factors in Eq.~(\ref{2})
are piecewise universal
in the limit $N\to\infty$ with $\zeta=Ns$ and $\mu=Nm$ fixed finite.

Due to the large-$N$ factorization 
of macroscopic correlation functions of
U($N$)-invariant operators
${\cal O}, {\cal O}'$ \cite{tHo},
\beq
\langle {\cal O}{\cal O}' \rangle=
\langle {\cal O} \rangle
\langle {\cal O}'\rangle + O\left(\frac{1}{N^2}\right),
\eeq
the first factor of (\ref{2}) is, 
in the large-$N$ limit, approximated by
\bea
&\sim&\e^{
-N^2 
\left\langle \frac1N {\rm tr}\,(V(H+s^2)-V(H))
\right\rangle_{m'}
}
\nonumber\\
&\sim&
\e^{-N^2 
\left(
s^2
\left\langle \frac1N {\rm tr}\,V'(H)
\right\rangle+O(s^4)\right)}.
\eea
Since the fermion determinant does not contribute to
the macroscopic correlator in the large-$N$ limit, we have dropped the 
suffix $m'$.

Now we change the picture back to $M$,
whose macroscopic spectral density is
$\rho(z) \equiv \langle
(1/2N) {\rm tr}\,\delta (z-M)\rangle =R(z^2 )\sqrt{a^2-z^2} $, 
where $[-a,a]$ is the support of the spectrum of $M$ and $R(z^2)$ is 
an analytic function which depends on the details of $v(M)$
\cite{ADMN}.  We obtain 
\bea
Q&\equiv&
\left\langle
\frac1N
{\rm tr}\, V'(H)
\right\rangle
\nonumber\\
&=&
\left\langle
\frac{1}{2N}
{\rm tr}\, \frac{v'(M)}{M}
\right\rangle
=
\int_{-a}^a
dz\,\rho(z)
\frac{v'(z)}{z}.
\label{vzz}
\eea
\noindent
In terms of $\rho(z)$, 
the leading (of order $O(N^2)$) part 
of the action in (\ref{Z}) is written as 
\beq
S=\int_{-a}^a dz\,\rho(z) 2v(z)-
\int_{-a}^a dz dw\,\rho(z)\rho(w) 2 P\ln|z-w|.
\eeq
The second term is the exponentiated Vandermonde determinant.
By substituting the 
large-$N$ saddle-point equation \cite{BIPZ}
$\delta S/\delta \rho(z)=0$, i.e., 
\beq
{v'(z)}
-2  
\int_{-a}^a dw\,\rho(w) 
P\frac{1}{z-w} =0,
\eeq
into (\ref{vzz}), we have
\beq
Q=
2
\int_{-a}^a
dz\,dw\,\rho(z)\rho(w)
P\frac{1}{z}\,P\frac{1}{z-w}.
\eeq
Using the identity 
$1/({z\pm i\epsilon})=P(1/z) \mp i\pi\delta(z)$
and taking into account that $\rho(z)$ is even, we 
finally
obtain
\beq
Q
=
\int^a_{-a}
dz\,dw\,\rho(z)\rho(w)
\pi\delta(z) 
\pi\delta(z-w) 
=
(\pi \rho(0))^2.
\eeq
That is,
\beq
\left.
\left\langle
\e^{-N
 {\rm tr}\,(V(H+s^2)-V(H))
}
\right\rangle_{s,m'}
\right|_{s=\frac{\zeta}{N}}
\stackrel{N\to \infty
}{\longrightarrow}
\e^{-(\pi\rho(0)\zeta)^2}
\eeq
which, after the rescaling $\zeta\rightarrow\zeta/(2\pi\rho(0))$,
universally reads $\exp(-\zeta^2/4)$.  This answers the question raised
in the introduction how the Gaussian factor arises from a generic
potential. 

We denote the microscopic limit of the partition function 
$Z^{(\a)}(\{m\}=\{ \frac{\mu}{N} \})$ by
${\cal Z}^{(\a)}(\{\mu\})$.
It is related to the microscopic kernel by the master formula 
\cite{D1}
\widetext
\Lrule
\beq
K^{(\a)}(\zeta_1,\zeta_2;\mu_1,\ldots,\mu_\a)
=
C\sqrt{|\zeta_1\zeta_2|}
\prod_f\sqrt{(\zeta_1^2+\mu_f^2)(\zeta_2^2+\mu_f^2)}
\frac{
{\cal Z}^{(\a+2)}(i\zeta_1,i\zeta_2,\mu_1,\ldots,\mu_\a )
}{
{\cal Z}^{(\a)}(\mu_1,\ldots,\mu_\a )
} ,
\label{master}
\eeq
where $C$ is a normalization constant. 
This formula is reminiscent of the very definition of 
the partition function and the kernel \cite{ZJ} and
is valid in the large-$N$ limit,
regardless of whether microscopic or macroscopic eigenvalue
and mass variables are kept finite.

Using the technique of orthogonal polynomials and rescaling 
$\zeta\to\zeta/(2\pi\rho(0)), \mu\to\mu/(2\pi\rho(0))$,
the left-hand side is shown to be universally given by \cite{DN1}
\bea
&&K^{(\alpha)}(\zeta_1,\zeta_2;\{\mu\})=
C\frac{
\sqrt{|\zeta_1\,\zeta_2|}
}{\zeta_1^2-\zeta_2^2}
\frac{
\det\limits_{1\leq i,j\leq \a+2} B_{ij}(\zeta_1,\zeta_2;\{\mu\})
}{
\prod_f
\sqrt{
\left( {\zeta_1^2} + {{{{\mu }_f}}^2} \right) \,
( {\zeta_2^2} + {{{{\mu }_f}}^2} ) 
}\,
\det\limits_{1\leq i,j\leq \a} A_{ij}(\{\mu\}) } , 
\label{Ks}\\
&&
~A_{ij}=
\mu_i^{j-1} I_{j-1}(\mu_i) ,
\ \ \ 
B_{ij}=
\left\{
\begin{array}{ll}
A_{ij} & ~~(i=1,\ldots,\a) \\
(-\zeta_{1,2})^{j-1} 
J_{j-1}(\zeta_{1,2}) & ~~(i=\a+1,\a+2)
\end{array}
\right.  ,
\nonumber
\eea
\Rrule
\narrowtext
\noindent
where $J$ and $I$ denote the Bessel function of real and imaginary
argument, respectively.
Therefore, after continuing $\zeta\to -i\mu$, we can derive
$
{\cal Z}^{(2n)}(\{\mu\} )
$
iteratively using (\ref{master}), starting from 
${\cal Z}^{(0)}=1$.
$
{\cal Z}^{(2n-1)}(\{\mu\} )
$
can be obtained by decoupling one of the masses in
${\cal Z}^{(2n)}(\{\mu\})$
by sending it to infinity.  In this way, one obtains universally
\beq
{\cal Z}^{(\a)}(\{\mu\} )
=
\frac{
\det\limits_{1\leq i,j\leq \a} A_{ij}(\{\mu\})
}{
\det\limits_{1\leq i,j\leq \a}
\mu_i^{2(j-1)}
} ,
\label{ZQCD}
\eeq
which is identical to the finite-volume QCD partition function as
calculated  from the chiral Lagrangian 
in the ``mesoscopic'' scaling limit \cite{LS,JSV}.

Both factors in (\ref{2}) being universal, the normalized hole probability
\beq
E^{(\a)}(\zeta;\{\mu\})=
\e^{-\zeta^2/4}\frac{{\cal Z}^{(\a)}(\{
\sqrt{\mu^2+\zeta^2}
\} )}{{\cal Z}^{(\a)}(\{\mu\})}
\label{Ez}
\eeq
as well as
the smallest eigenvalue distribution \cite{WGW}
\bea
&&
P^{(\a)}(\zeta;\{\mu\})=-\frac{\partial}{\partial \zeta} 
E^{(\a)}(\zeta;\{\mu\})
\nonumber
\\
&&=
\frac{\zeta}{2}\e^{-\zeta^2/4}
\frac{
\det\limits_{1\leq i,j\leq \a} 
C_{ij}(\{\sqrt{\mu^2+\zeta^2}\})
}{
\det\limits_{1\leq i,j\leq \a} A_{ij}(\{\mu\})
} 
\label{Pz}
\eea
with $C_{ij}(\{\mu\})=\mu_i^{j-1} I_{j+1}(\mu_i)$ are
universal.  As explained above, the general case of nonzero
topological charge $\nu$ is obtained by introducing $|\nu|$ massless
flavors ($\alpha=N_f+|\nu|$).

Except for the Gaussian prefactor the expressions (\ref{Ez}) and
(\ref{Pz}) are explicitly given in terms of a finite-volume field
theory partition function in the mesoscopic scaling regime. This
indicates that also these quantities can be derived directly from
field theory in the mesoscopic scaling regime, without the bypass
through random matrix theory \cite{D1}.

The hole probabilities for the
orthogonal ($\beta=1$) and symplectic ($\beta=4$) ensembles
are also related to the unitary kernel $K$ by formulae
analogous to Eq.~(\ref{Es}) 
($\det_\pm$ represents the determinant
projected to even or odd orders of eigenvalues)
\cite{Meh}:
\begin{mathletters}\bea
E_1(s)&=&\det\nolimits_+ (1-\hat{K}),\\
E_4(s)&=&\frac12 \left(
\det\nolimits_+ (1-\hat{K}) + (\det\nolimits_- (1-\hat{K}) \right),
\eea\end{mathletters}
which hold regardless of the fermion determinant.
Hence, the smallest eigenvalue
distributions for these ensembles are also guaranteed to be universal.
This is consistent with the recently proven universality \cite{SV} 
of the microscopic kernels for 
the massless chiral orthogonal and symplectic ensembles \cite{NF1}
via universal relationships between them and
the massless chiral unitary kernel.
Explicit expressions for $E_1(s)$ and $E_4(s)$
are however not known except for the massless
chiral Gaussian/Laguerre case \cite{NF2}. 
To obtain these quantities for the corresponding massive ensembles, 
our method presented here needs modification because the mapping
$M\mapsto H$ involves a nontrivial Jacobian, $(\det H)^{\beta/2-1}$.
This point will be discussed elsewhere.

TW would like to thank T. Guhr and T. Wilke for useful discussions.

\widetext
\end{document}